\documentstyle[12pt,epsf]{article}
\begin{document}
\begin{flushright}
Southampton Preprint No.: SHEP 94/95-01
\end{flushright}
\begin{flushleft}
\hspace{1cm}\\
\vspace{8\baselineskip}
{\bf SUPERSYMMETRIC GRAND UNIFIED THEORIES AND YUKAWA UNIFICATION\\}
\vspace{3\baselineskip}
\hspace{1in}B. C. Allanach\\
\vspace{\baselineskip}
\hspace{1in}Physics Department\\
\hspace{1in}University of Southampton\\
\hspace{1in}Southampton\\
\hspace{1in}SO9 5NH\\
\hspace{1in}UK\\
\vspace{3\baselineskip}
{\bf INTRODUCTION}
\end{flushleft}

In this paper, I intend to motivate supersymmetric grand unified
theories (SUSY GUTs), briefly explain an extension of the standard
model based on them and present a calculation performed using
certain properties of some SUSY GUTs to constrain the available
parameter
space.
\begin{flushleft}
{\bf Why GUTs?}
\end{flushleft}

Much work has been done on the running of the gauge couplings in the
standard model, as prescribed by the renormalisation group. Amazingly,
when the couplings $\alpha_1$, $\alpha_2$ and $\alpha_3$ were run up
to fantastically high energies $\sim 10^{14}$ GeV, they
seemed to
be converging$^1$ to one value. This is a feature naturally
explained by many GUTs such as SU(5)$^{2,3}$ and
reflects the fact that the strong, weak and electromagnetic forces seen
today are different parts of the same grand unified force. It was
realised that GUTs could also provide relations between the masses of
the observed fermions, the structure and hierarchy of which are as yet
unexplained. Despite
these attractive features, several problems arose which detracted from
the idea.

Unfortunately, the three couplings do not quite converge by $\sim
7 \sigma$, and many GUTs, notably SU(5), predict proton
decay much faster than the lower experimental bounds. Also incredible
fine tuning is required for the so-called 'hierarchy problem'.
This
stems from the fact that $M_W$ changes through radiative
corrections of order the new physics scale (Fig.1)
, say the Planck mass $\sim
10^{19}$ GeV, if there is no new physics at smaller energies.
 $M_W$ is therefore unstable
to the corrections and vast cancellations in the couplings are
required to motivate the correct phenomenology.
\begin{figure}
\begin{center}
\begin{displaymath}
\begin{array}{c}

\setlength{\unitlength}{0.25cm}
\begin{picture}( 10.000, 10.000)(   .000,   .000)
{\special{em:linewidth    .020cm}}
\put(  2.000,  5.180){\makebox(0,0)[b]{\tiny H}}
\put(  8.000,  5.180){\makebox(0,0)[b]{\tiny H}}
\put(  5.000,  6.680){\makebox(0,0)[b]{\tiny F}}
\put(  5.000,  3.320){\makebox(0,0)[t]{\tiny F}}
\put(   .500,  5.000){\special{em:moveto}}
\put(   .583,  5.000){\special{em:lineto}}
\put(   .750,  5.000){\special{em:moveto}}
\put(   .917,  5.000){\special{em:lineto}}
\put(  1.083,  5.000){\special{em:moveto}}
\put(  1.250,  5.000){\special{em:lineto}}
\put(  1.417,  5.000){\special{em:moveto}}
\put(  1.583,  5.000){\special{em:lineto}}
\put(  1.750,  5.000){\special{em:moveto}}
\put(  1.917,  5.000){\special{em:lineto}}
\put(  2.083,  5.000){\special{em:moveto}}
\put(  2.250,  5.000){\special{em:lineto}}
\put(  2.417,  5.000){\special{em:moveto}}
\put(  2.583,  5.000){\special{em:lineto}}
\put(  2.750,  5.000){\special{em:moveto}}
\put(  2.917,  5.000){\special{em:lineto}}
\put(  3.083,  5.000){\special{em:moveto}}
\put(  3.250,  5.000){\special{em:lineto}}
\put(  3.417,  5.000){\special{em:moveto}}
\put(  3.500,  5.000){\special{em:lineto}}
\put(  6.500,  5.000){\special{em:moveto}}
\put(  6.583,  5.000){\special{em:lineto}}
\put(  6.750,  5.000){\special{em:moveto}}
\put(  6.917,  5.000){\special{em:lineto}}
\put(  7.083,  5.000){\special{em:moveto}}
\put(  7.250,  5.000){\special{em:lineto}}
\put(  7.417,  5.000){\special{em:moveto}}
\put(  7.583,  5.000){\special{em:lineto}}
\put(  7.750,  5.000){\special{em:moveto}}
\put(  7.917,  5.000){\special{em:lineto}}
\put(  8.083,  5.000){\special{em:moveto}}
\put(  8.250,  5.000){\special{em:lineto}}
\put(  8.417,  5.000){\special{em:moveto}}
\put(  8.583,  5.000){\special{em:lineto}}
\put(  8.750,  5.000){\special{em:moveto}}
\put(  8.917,  5.000){\special{em:lineto}}
\put(  9.083,  5.000){\special{em:moveto}}
\put(  9.250,  5.000){\special{em:lineto}}
\put(  9.417,  5.000){\special{em:moveto}}
\put(  9.500,  5.000){\special{em:lineto}}
\put(  3.500,  5.000){\special{em:moveto}}
\put(  3.512,  4.812){\special{em:lineto}}
\put(  3.547,  4.627){\special{em:lineto}}
\put(  3.605,  4.448){\special{em:lineto}}
\put(  3.686,  4.277){\special{em:lineto}}
\put(  3.786,  4.118){\special{em:lineto}}
\put(  3.907,  3.973){\special{em:lineto}}
\put(  4.044,  3.844){\special{em:lineto}}
\put(  4.196,  3.734){\special{em:lineto}}
\put(  4.361,  3.643){\special{em:lineto}}
\put(  4.536,  3.573){\special{em:lineto}}
\put(  4.719,  3.527){\special{em:lineto}}
\put(  4.906,  3.503){\special{em:lineto}}
\put(  5.094,  3.503){\special{em:lineto}}
\put(  5.281,  3.527){\special{em:lineto}}
\put(  5.464,  3.573){\special{em:lineto}}
\put(  5.639,  3.643){\special{em:lineto}}
\put(  5.804,  3.734){\special{em:lineto}}
\put(  5.956,  3.844){\special{em:lineto}}
\put(  6.093,  3.973){\special{em:lineto}}
\put(  6.214,  4.118){\special{em:lineto}}
\put(  6.314,  4.277){\special{em:lineto}}
\put(  6.395,  4.448){\special{em:lineto}}
\put(  6.453,  4.627){\special{em:lineto}}
\put(  6.488,  4.812){\special{em:lineto}}
\put(  6.500,  5.000){\special{em:lineto}}
\put(  4.805,  3.603){\special{em:moveto}}
\put(  5.195,  3.513){\special{em:lineto}}
\put(  4.805,  3.423){\special{em:lineto}}
\put(  4.805,  3.603){\special{em:lineto}}
\put(  6.500,  5.000){\special{em:moveto}}
\put(  6.488,  5.188){\special{em:lineto}}
\put(  6.453,  5.373){\special{em:lineto}}
\put(  6.395,  5.552){\special{em:lineto}}
\put(  6.314,  5.723){\special{em:lineto}}
\put(  6.214,  5.882){\special{em:lineto}}
\put(  6.093,  6.027){\special{em:lineto}}
\put(  5.956,  6.156){\special{em:lineto}}
\put(  5.804,  6.266){\special{em:lineto}}
\put(  5.639,  6.357){\special{em:lineto}}
\put(  5.464,  6.427){\special{em:lineto}}
\put(  5.281,  6.473){\special{em:lineto}}
\put(  5.094,  6.497){\special{em:lineto}}
\put(  4.906,  6.497){\special{em:lineto}}
\put(  4.719,  6.473){\special{em:lineto}}
\put(  4.536,  6.427){\special{em:lineto}}
\put(  4.361,  6.357){\special{em:lineto}}
\put(  4.196,  6.266){\special{em:lineto}}
\put(  4.044,  6.156){\special{em:lineto}}
\put(  3.907,  6.027){\special{em:lineto}}
\put(  3.786,  5.882){\special{em:lineto}}
\put(  3.686,  5.723){\special{em:lineto}}
\put(  3.605,  5.552){\special{em:lineto}}
\put(  3.547,  5.373){\special{em:lineto}}
\put(  3.512,  5.188){\special{em:lineto}}
\put(  3.500,  5.000){\special{em:lineto}}
\put(  5.195,  6.397){\special{em:moveto}}
\put(  4.805,  6.487){\special{em:lineto}}
\put(  5.195,  6.577){\special{em:lineto}}
\put(  5.195,  6.397){\special{em:lineto}}
\end{picture}
\end{array}+
\begin{array}{c}
\setlength{\unitlength}{0.25cm}
\begin{picture}( 10.000, 10.000)(   .000,   .000)
{\special{em:linewidth    .020cm}}
\put(  5.000,  7.820){\makebox(0,0)[t]{\tiny W}}
\put(  2.250,  5.180){\makebox(0,0)[b]{\tiny H}}
\put(  7.750,  5.180){\makebox(0,0)[b]{\tiny H}}
\put(   .500,  5.000){\special{em:moveto}}
\put(   .590,  5.000){\special{em:lineto}}
\put(   .770,  5.000){\special{em:moveto}}
\put(   .950,  5.000){\special{em:lineto}}
\put(  1.130,  5.000){\special{em:moveto}}
\put(  1.310,  5.000){\special{em:lineto}}
\put(  1.490,  5.000){\special{em:moveto}}
\put(  1.670,  5.000){\special{em:lineto}}
\put(  1.850,  5.000){\special{em:moveto}}
\put(  2.030,  5.000){\special{em:lineto}}
\put(  2.210,  5.000){\special{em:moveto}}
\put(  2.390,  5.000){\special{em:lineto}}
\put(  2.570,  5.000){\special{em:moveto}}
\put(  2.750,  5.000){\special{em:lineto}}
\put(  2.930,  5.000){\special{em:moveto}}
\put(  3.110,  5.000){\special{em:lineto}}
\put(  3.290,  5.000){\special{em:moveto}}
\put(  3.470,  5.000){\special{em:lineto}}
\put(  3.650,  5.000){\special{em:moveto}}
\put(  3.830,  5.000){\special{em:lineto}}
\put(  4.010,  5.000){\special{em:moveto}}
\put(  4.190,  5.000){\special{em:lineto}}
\put(  4.370,  5.000){\special{em:moveto}}
\put(  4.550,  5.000){\special{em:lineto}}
\put(  4.730,  5.000){\special{em:moveto}}
\put(  4.910,  5.000){\special{em:lineto}}
\put(  5.090,  5.000){\special{em:moveto}}
\put(  5.270,  5.000){\special{em:lineto}}
\put(  5.450,  5.000){\special{em:moveto}}
\put(  5.630,  5.000){\special{em:lineto}}
\put(  5.810,  5.000){\special{em:moveto}}
\put(  5.990,  5.000){\special{em:lineto}}
\put(  6.170,  5.000){\special{em:moveto}}
\put(  6.350,  5.000){\special{em:lineto}}
\put(  6.530,  5.000){\special{em:moveto}}
\put(  6.710,  5.000){\special{em:lineto}}
\put(  6.890,  5.000){\special{em:moveto}}
\put(  7.070,  5.000){\special{em:lineto}}
\put(  7.250,  5.000){\special{em:moveto}}
\put(  7.430,  5.000){\special{em:lineto}}
\put(  7.610,  5.000){\special{em:moveto}}
\put(  7.790,  5.000){\special{em:lineto}}
\put(  7.970,  5.000){\special{em:moveto}}
\put(  8.150,  5.000){\special{em:lineto}}
\put(  8.330,  5.000){\special{em:moveto}}
\put(  8.510,  5.000){\special{em:lineto}}
\put(  8.690,  5.000){\special{em:moveto}}
\put(  8.870,  5.000){\special{em:lineto}}
\put(  9.050,  5.000){\special{em:moveto}}
\put(  9.230,  5.000){\special{em:lineto}}
\put(  9.410,  5.000){\special{em:moveto}}
\put(  9.500,  5.000){\special{em:lineto}}
\put(  5.000,  5.000){\special{em:moveto}}
\put(  5.062,  4.913){\special{em:lineto}}
\put(  5.129,  4.862){\special{em:lineto}}
\put(  5.193,  4.869){\special{em:lineto}}
\put(  5.248,  4.931){\special{em:lineto}}
\put(  5.293,  5.029){\special{em:lineto}}
\put(  5.330,  5.127){\special{em:lineto}}
\put(  5.368,  5.194){\special{em:lineto}}
\put(  5.419,  5.209){\special{em:lineto}}
\put(  5.489,  5.175){\special{em:lineto}}
\put(  5.574,  5.114){\special{em:lineto}}
\put(  5.665,  5.058){\special{em:lineto}}
\put(  5.746,  5.036){\special{em:lineto}}
\put(  5.803,  5.067){\special{em:lineto}}
\put(  5.830,  5.146){\special{em:lineto}}
\put(  5.833,  5.253){\special{em:lineto}}
\put(  5.830,  5.358){\special{em:lineto}}
\put(  5.840,  5.434){\special{em:lineto}}
\put(  5.882,  5.468){\special{em:lineto}}
\put(  5.958,  5.463){\special{em:lineto}}
\put(  6.061,  5.439){\special{em:lineto}}
\put(  6.166,  5.422){\special{em:lineto}}
\put(  6.249,  5.433){\special{em:lineto}}
\put(  6.290,  5.483){\special{em:lineto}}
\put(  6.285,  5.566){\special{em:lineto}}
\put(  6.247,  5.667){\special{em:lineto}}
\put(  6.204,  5.762){\special{em:lineto}}
\put(  6.184,  5.837){\special{em:lineto}}
\put(  6.209,  5.884){\special{em:lineto}}
\put(  6.282,  5.909){\special{em:lineto}}
\put(  6.386,  5.926){\special{em:lineto}}
\put(  6.490,  5.950){\special{em:lineto}}
\put(  6.562,  5.992){\special{em:lineto}}
\put(  6.581,  6.054){\special{em:lineto}}
\put(  6.544,  6.129){\special{em:lineto}}
\put(  6.471,  6.207){\special{em:lineto}}
\put(  6.394,  6.279){\special{em:lineto}}
\put(  6.348,  6.340){\special{em:lineto}}
\put(  6.353,  6.394){\special{em:lineto}}
\put(  6.411,  6.445){\special{em:lineto}}
\put(  6.500,  6.500){\special{em:lineto}}
\put(  6.587,  6.562){\special{em:lineto}}
\put(  6.638,  6.629){\special{em:lineto}}
\put(  6.631,  6.693){\special{em:lineto}}
\put(  6.569,  6.748){\special{em:lineto}}
\put(  6.471,  6.793){\special{em:lineto}}
\put(  6.373,  6.830){\special{em:lineto}}
\put(  6.306,  6.868){\special{em:lineto}}
\put(  6.291,  6.919){\special{em:lineto}}
\put(  6.325,  6.989){\special{em:lineto}}
\put(  6.386,  7.074){\special{em:lineto}}
\put(  6.442,  7.165){\special{em:lineto}}
\put(  6.464,  7.246){\special{em:lineto}}
\put(  6.433,  7.303){\special{em:lineto}}
\put(  6.354,  7.330){\special{em:lineto}}
\put(  6.247,  7.333){\special{em:lineto}}
\put(  6.142,  7.330){\special{em:lineto}}
\put(  6.066,  7.340){\special{em:lineto}}
\put(  6.032,  7.382){\special{em:lineto}}
\put(  6.037,  7.458){\special{em:lineto}}
\put(  6.061,  7.561){\special{em:lineto}}
\put(  6.078,  7.666){\special{em:lineto}}
\put(  6.067,  7.749){\special{em:lineto}}
\put(  6.017,  7.790){\special{em:lineto}}
\put(  5.934,  7.785){\special{em:lineto}}
\put(  5.833,  7.747){\special{em:lineto}}
\put(  5.738,  7.704){\special{em:lineto}}
\put(  5.663,  7.684){\special{em:lineto}}
\put(  5.616,  7.709){\special{em:lineto}}
\put(  5.591,  7.782){\special{em:lineto}}
\put(  5.574,  7.886){\special{em:lineto}}
\put(  5.550,  7.990){\special{em:lineto}}
\put(  5.508,  8.062){\special{em:lineto}}
\put(  5.446,  8.081){\special{em:lineto}}
\put(  5.371,  8.044){\special{em:lineto}}
\put(  5.293,  7.971){\special{em:lineto}}
\put(  5.221,  7.894){\special{em:lineto}}
\put(  5.160,  7.848){\special{em:lineto}}
\put(  5.106,  7.853){\special{em:lineto}}
\put(  5.055,  7.911){\special{em:lineto}}
\put(  5.000,  8.000){\special{em:lineto}}
\put(  5.000,  8.000){\special{em:moveto}}
\put(  4.938,  8.087){\special{em:lineto}}
\put(  4.871,  8.138){\special{em:lineto}}
\put(  4.807,  8.131){\special{em:lineto}}
\put(  4.752,  8.069){\special{em:lineto}}
\put(  4.707,  7.971){\special{em:lineto}}
\put(  4.670,  7.873){\special{em:lineto}}
\put(  4.632,  7.806){\special{em:lineto}}
\put(  4.581,  7.791){\special{em:lineto}}
\put(  4.511,  7.825){\special{em:lineto}}
\put(  4.426,  7.886){\special{em:lineto}}
\put(  4.335,  7.942){\special{em:lineto}}
\put(  4.254,  7.964){\special{em:lineto}}
\put(  4.197,  7.933){\special{em:lineto}}
\put(  4.170,  7.854){\special{em:lineto}}
\put(  4.167,  7.747){\special{em:lineto}}
\put(  4.170,  7.642){\special{em:lineto}}
\put(  4.160,  7.566){\special{em:lineto}}
\put(  4.118,  7.532){\special{em:lineto}}
\put(  4.042,  7.537){\special{em:lineto}}
\put(  3.939,  7.561){\special{em:lineto}}
\put(  3.834,  7.578){\special{em:lineto}}
\put(  3.751,  7.567){\special{em:lineto}}
\put(  3.710,  7.517){\special{em:lineto}}
\put(  3.715,  7.434){\special{em:lineto}}
\put(  3.753,  7.333){\special{em:lineto}}
\put(  3.796,  7.238){\special{em:lineto}}
\put(  3.816,  7.163){\special{em:lineto}}
\put(  3.791,  7.116){\special{em:lineto}}
\put(  3.718,  7.091){\special{em:lineto}}
\put(  3.614,  7.074){\special{em:lineto}}
\put(  3.510,  7.050){\special{em:lineto}}
\put(  3.438,  7.008){\special{em:lineto}}
\put(  3.419,  6.946){\special{em:lineto}}
\put(  3.456,  6.871){\special{em:lineto}}
\put(  3.529,  6.793){\special{em:lineto}}
\put(  3.606,  6.721){\special{em:lineto}}
\put(  3.652,  6.660){\special{em:lineto}}
\put(  3.647,  6.606){\special{em:lineto}}
\put(  3.589,  6.555){\special{em:lineto}}
\put(  3.500,  6.500){\special{em:lineto}}
\put(  3.413,  6.438){\special{em:lineto}}
\put(  3.362,  6.371){\special{em:lineto}}
\put(  3.369,  6.307){\special{em:lineto}}
\put(  3.431,  6.252){\special{em:lineto}}
\put(  3.529,  6.207){\special{em:lineto}}
\put(  3.627,  6.170){\special{em:lineto}}
\put(  3.694,  6.132){\special{em:lineto}}
\put(  3.709,  6.081){\special{em:lineto}}
\put(  3.675,  6.011){\special{em:lineto}}
\put(  3.614,  5.926){\special{em:lineto}}
\put(  3.558,  5.835){\special{em:lineto}}
\put(  3.536,  5.754){\special{em:lineto}}
\put(  3.567,  5.697){\special{em:lineto}}
\put(  3.646,  5.670){\special{em:lineto}}
\put(  3.753,  5.667){\special{em:lineto}}
\put(  3.858,  5.670){\special{em:lineto}}
\put(  3.934,  5.660){\special{em:lineto}}
\put(  3.968,  5.618){\special{em:lineto}}
\put(  3.963,  5.542){\special{em:lineto}}
\put(  3.939,  5.439){\special{em:lineto}}
\put(  3.922,  5.334){\special{em:lineto}}
\put(  3.933,  5.251){\special{em:lineto}}
\put(  3.983,  5.210){\special{em:lineto}}
\put(  4.066,  5.215){\special{em:lineto}}
\put(  4.167,  5.253){\special{em:lineto}}
\put(  4.262,  5.296){\special{em:lineto}}
\put(  4.337,  5.316){\special{em:lineto}}
\put(  4.384,  5.291){\special{em:lineto}}
\put(  4.409,  5.218){\special{em:lineto}}
\put(  4.426,  5.114){\special{em:lineto}}
\put(  4.450,  5.010){\special{em:lineto}}
\put(  4.492,  4.938){\special{em:lineto}}
\put(  4.554,  4.919){\special{em:lineto}}
\put(  4.629,  4.956){\special{em:lineto}}
\put(  4.707,  5.029){\special{em:lineto}}
\put(  4.779,  5.106){\special{em:lineto}}
\put(  4.840,  5.152){\special{em:lineto}}
\put(  4.894,  5.147){\special{em:lineto}}
\put(  4.945,  5.089){\special{em:lineto}}
\put(  5.000,  5.000){\special{em:lineto}}
\end{picture}
\end{array}+
\begin{array}{c}
\setlength{\unitlength}{0.25cm}
\begin{picture}( 10.000, 10.000)(   .000,   .000)
{\special{em:linewidth    .020cm}}
\put(  5.000,  7.820){\makebox(0,0)[t]{\tiny H}}
\put(  2.250,  5.180){\makebox(0,0)[b]{\tiny H}}
\put(  7.750,  5.180){\makebox(0,0)[b]{\tiny H}}
\put(   .500,  5.000){\special{em:moveto}}
\put(   .590,  5.000){\special{em:lineto}}
\put(   .770,  5.000){\special{em:moveto}}
\put(   .950,  5.000){\special{em:lineto}}
\put(  1.130,  5.000){\special{em:moveto}}
\put(  1.310,  5.000){\special{em:lineto}}
\put(  1.490,  5.000){\special{em:moveto}}
\put(  1.670,  5.000){\special{em:lineto}}
\put(  1.850,  5.000){\special{em:moveto}}
\put(  2.030,  5.000){\special{em:lineto}}
\put(  2.210,  5.000){\special{em:moveto}}
\put(  2.390,  5.000){\special{em:lineto}}
\put(  2.570,  5.000){\special{em:moveto}}
\put(  2.750,  5.000){\special{em:lineto}}
\put(  2.930,  5.000){\special{em:moveto}}
\put(  3.110,  5.000){\special{em:lineto}}
\put(  3.290,  5.000){\special{em:moveto}}
\put(  3.470,  5.000){\special{em:lineto}}
\put(  3.650,  5.000){\special{em:moveto}}
\put(  3.830,  5.000){\special{em:lineto}}
\put(  4.010,  5.000){\special{em:moveto}}
\put(  4.190,  5.000){\special{em:lineto}}
\put(  4.370,  5.000){\special{em:moveto}}
\put(  4.550,  5.000){\special{em:lineto}}
\put(  4.730,  5.000){\special{em:moveto}}
\put(  4.910,  5.000){\special{em:lineto}}
\put(  5.090,  5.000){\special{em:moveto}}
\put(  5.270,  5.000){\special{em:lineto}}
\put(  5.450,  5.000){\special{em:moveto}}
\put(  5.630,  5.000){\special{em:lineto}}
\put(  5.810,  5.000){\special{em:moveto}}
\put(  5.990,  5.000){\special{em:lineto}}
\put(  6.170,  5.000){\special{em:moveto}}
\put(  6.350,  5.000){\special{em:lineto}}
\put(  6.530,  5.000){\special{em:moveto}}
\put(  6.710,  5.000){\special{em:lineto}}
\put(  6.890,  5.000){\special{em:moveto}}
\put(  7.070,  5.000){\special{em:lineto}}
\put(  7.250,  5.000){\special{em:moveto}}
\put(  7.430,  5.000){\special{em:lineto}}
\put(  7.610,  5.000){\special{em:moveto}}
\put(  7.790,  5.000){\special{em:lineto}}
\put(  7.970,  5.000){\special{em:moveto}}
\put(  8.150,  5.000){\special{em:lineto}}
\put(  8.330,  5.000){\special{em:moveto}}
\put(  8.510,  5.000){\special{em:lineto}}
\put(  8.690,  5.000){\special{em:moveto}}
\put(  8.870,  5.000){\special{em:lineto}}
\put(  9.050,  5.000){\special{em:moveto}}
\put(  9.230,  5.000){\special{em:lineto}}
\put(  9.410,  5.000){\special{em:moveto}}
\put(  9.500,  5.000){\special{em:lineto}}
\put(  5.000,  5.000){\special{em:moveto}}
\put(  5.084,  5.002){\special{em:lineto}}
\put(  5.251,  5.021){\special{em:moveto}}
\put(  5.415,  5.059){\special{em:lineto}}
\put(  5.574,  5.114){\special{em:moveto}}
\put(  5.726,  5.187){\special{em:lineto}}
\put(  5.868,  5.277){\special{em:moveto}}
\put(  6.000,  5.382){\special{em:lineto}}
\put(  6.118,  5.500){\special{em:moveto}}
\put(  6.223,  5.632){\special{em:lineto}}
\put(  6.313,  5.774){\special{em:moveto}}
\put(  6.386,  5.926){\special{em:lineto}}
\put(  6.441,  6.085){\special{em:moveto}}
\put(  6.479,  6.249){\special{em:lineto}}
\put(  6.498,  6.416){\special{em:moveto}}
\put(  6.498,  6.584){\special{em:lineto}}
\put(  6.479,  6.751){\special{em:moveto}}
\put(  6.441,  6.915){\special{em:lineto}}
\put(  6.386,  7.074){\special{em:moveto}}
\put(  6.313,  7.226){\special{em:lineto}}
\put(  6.223,  7.368){\special{em:moveto}}
\put(  6.118,  7.500){\special{em:lineto}}
\put(  6.000,  7.618){\special{em:moveto}}
\put(  5.868,  7.723){\special{em:lineto}}
\put(  5.726,  7.813){\special{em:moveto}}
\put(  5.574,  7.886){\special{em:lineto}}
\put(  5.415,  7.941){\special{em:moveto}}
\put(  5.251,  7.979){\special{em:lineto}}
\put(  5.084,  7.998){\special{em:moveto}}
\put(  5.000,  8.000){\special{em:lineto}}
\put(  5.000,  8.000){\special{em:moveto}}
\put(  4.916,  7.998){\special{em:lineto}}
\put(  4.749,  7.979){\special{em:moveto}}
\put(  4.585,  7.941){\special{em:lineto}}
\put(  4.426,  7.886){\special{em:moveto}}
\put(  4.274,  7.813){\special{em:lineto}}
\put(  4.132,  7.723){\special{em:moveto}}
\put(  4.000,  7.618){\special{em:lineto}}
\put(  3.882,  7.500){\special{em:moveto}}
\put(  3.777,  7.368){\special{em:lineto}}
\put(  3.687,  7.226){\special{em:moveto}}
\put(  3.614,  7.074){\special{em:lineto}}
\put(  3.559,  6.915){\special{em:moveto}}
\put(  3.521,  6.751){\special{em:lineto}}
\put(  3.502,  6.584){\special{em:moveto}}
\put(  3.502,  6.416){\special{em:lineto}}
\put(  3.521,  6.249){\special{em:moveto}}
\put(  3.559,  6.085){\special{em:lineto}}
\put(  3.614,  5.926){\special{em:moveto}}
\put(  3.687,  5.774){\special{em:lineto}}
\put(  3.777,  5.632){\special{em:moveto}}
\put(  3.882,  5.500){\special{em:lineto}}
\put(  4.000,  5.382){\special{em:moveto}}
\put(  4.132,  5.277){\special{em:lineto}}
\put(  4.274,  5.187){\special{em:moveto}}
\put(  4.426,  5.114){\special{em:lineto}}
\put(  4.585,  5.059){\special{em:moveto}}
\put(  4.749,  5.021){\special{em:lineto}}
\put(  4.916,  5.002){\special{em:moveto}}
\put(  5.000,  5.000){\special{em:lineto}}
\end{picture}
\end{array}+
\begin{array}{c}
% Uncompressed output from FDIAG
% LaTeX / EmTeX form of gaugehig.fil
\setlength{\unitlength}{0.25cm}
\begin{picture}( 10.000, 10.000)(   .000,   .000)
{\special{em:linewidth    .020cm}}
\put(  2.000,  5.180){\makebox(0,0)[b]{\tiny H}}
\put(  8.000,  5.180){\makebox(0,0)[b]{\tiny H}}
\put(  5.000,  6.680){\makebox(0,0)[b]{\tiny W}}
\put(  5.000,  3.320){\makebox(0,0)[t]{\tiny H}}
\put(   .500,  5.000){\special{em:moveto}}
\put(   .583,  5.000){\special{em:lineto}}
\put(   .750,  5.000){\special{em:moveto}}
\put(   .917,  5.000){\special{em:lineto}}
\put(  1.083,  5.000){\special{em:moveto}}
\put(  1.250,  5.000){\special{em:lineto}}
\put(  1.417,  5.000){\special{em:moveto}}
\put(  1.583,  5.000){\special{em:lineto}}
\put(  1.750,  5.000){\special{em:moveto}}
\put(  1.917,  5.000){\special{em:lineto}}
\put(  2.083,  5.000){\special{em:moveto}}
\put(  2.250,  5.000){\special{em:lineto}}
\put(  2.417,  5.000){\special{em:moveto}}
\put(  2.583,  5.000){\special{em:lineto}}
\put(  2.750,  5.000){\special{em:moveto}}
\put(  2.917,  5.000){\special{em:lineto}}
\put(  3.083,  5.000){\special{em:moveto}}
\put(  3.250,  5.000){\special{em:lineto}}
\put(  3.417,  5.000){\special{em:moveto}}
\put(  3.500,  5.000){\special{em:lineto}}
\put(  6.500,  5.000){\special{em:moveto}}
\put(  6.583,  5.000){\special{em:lineto}}
\put(  6.750,  5.000){\special{em:moveto}}
\put(  6.917,  5.000){\special{em:lineto}}
\put(  7.083,  5.000){\special{em:moveto}}
\put(  7.250,  5.000){\special{em:lineto}}
\put(  7.417,  5.000){\special{em:moveto}}
\put(  7.583,  5.000){\special{em:lineto}}
\put(  7.750,  5.000){\special{em:moveto}}
\put(  7.917,  5.000){\special{em:lineto}}
\put(  8.083,  5.000){\special{em:moveto}}
\put(  8.250,  5.000){\special{em:lineto}}
\put(  8.417,  5.000){\special{em:moveto}}
\put(  8.583,  5.000){\special{em:lineto}}
\put(  8.750,  5.000){\special{em:moveto}}
\put(  8.917,  5.000){\special{em:lineto}}
\put(  9.083,  5.000){\special{em:moveto}}
\put(  9.250,  5.000){\special{em:lineto}}
\put(  9.417,  5.000){\special{em:moveto}}
\put(  9.500,  5.000){\special{em:lineto}}
\put(  3.500,  5.000){\special{em:moveto}}
\put(  3.502,  4.916){\special{em:lineto}}
\put(  3.521,  4.749){\special{em:moveto}}
\put(  3.559,  4.585){\special{em:lineto}}
\put(  3.614,  4.426){\special{em:moveto}}
\put(  3.687,  4.274){\special{em:lineto}}
\put(  3.777,  4.132){\special{em:moveto}}
\put(  3.882,  4.000){\special{em:lineto}}
\put(  4.000,  3.882){\special{em:moveto}}
\put(  4.132,  3.777){\special{em:lineto}}
\put(  4.274,  3.687){\special{em:moveto}}
\put(  4.426,  3.614){\special{em:lineto}}
\put(  4.585,  3.559){\special{em:moveto}}
\put(  4.749,  3.521){\special{em:lineto}}
\put(  4.916,  3.502){\special{em:moveto}}
\put(  5.084,  3.502){\special{em:lineto}}
\put(  5.251,  3.521){\special{em:moveto}}
\put(  5.415,  3.559){\special{em:lineto}}
\put(  5.574,  3.614){\special{em:moveto}}
\put(  5.726,  3.687){\special{em:lineto}}
\put(  5.868,  3.777){\special{em:moveto}}
\put(  6.000,  3.882){\special{em:lineto}}
\put(  6.118,  4.000){\special{em:moveto}}
\put(  6.223,  4.132){\special{em:lineto}}
\put(  6.313,  4.274){\special{em:moveto}}
\put(  6.386,  4.426){\special{em:lineto}}
\put(  6.441,  4.585){\special{em:moveto}}
\put(  6.479,  4.749){\special{em:lineto}}
\put(  6.498,  4.916){\special{em:moveto}}
\put(  6.500,  5.000){\special{em:lineto}}
\put(  6.500,  5.000){\special{em:moveto}}
\put(  6.587,  5.062){\special{em:lineto}}
\put(  6.638,  5.129){\special{em:lineto}}
\put(  6.631,  5.193){\special{em:lineto}}
\put(  6.569,  5.248){\special{em:lineto}}
\put(  6.471,  5.293){\special{em:lineto}}
\put(  6.373,  5.330){\special{em:lineto}}
\put(  6.306,  5.368){\special{em:lineto}}
\put(  6.291,  5.419){\special{em:lineto}}
\put(  6.325,  5.489){\special{em:lineto}}
\put(  6.386,  5.574){\special{em:lineto}}
\put(  6.442,  5.665){\special{em:lineto}}
\put(  6.464,  5.746){\special{em:lineto}}
\put(  6.433,  5.803){\special{em:lineto}}
\put(  6.354,  5.830){\special{em:lineto}}
\put(  6.247,  5.833){\special{em:lineto}}
\put(  6.142,  5.830){\special{em:lineto}}
\put(  6.066,  5.840){\special{em:lineto}}
\put(  6.032,  5.882){\special{em:lineto}}
\put(  6.037,  5.958){\special{em:lineto}}
\put(  6.061,  6.061){\special{em:lineto}}
\put(  6.078,  6.166){\special{em:lineto}}
\put(  6.067,  6.249){\special{em:lineto}}
\put(  6.017,  6.290){\special{em:lineto}}
\put(  5.934,  6.285){\special{em:lineto}}
\put(  5.833,  6.247){\special{em:lineto}}
\put(  5.738,  6.204){\special{em:lineto}}
\put(  5.663,  6.184){\special{em:lineto}}
\put(  5.616,  6.209){\special{em:lineto}}
\put(  5.591,  6.282){\special{em:lineto}}
\put(  5.574,  6.386){\special{em:lineto}}
\put(  5.550,  6.490){\special{em:lineto}}
\put(  5.508,  6.562){\special{em:lineto}}
\put(  5.446,  6.581){\special{em:lineto}}
\put(  5.371,  6.544){\special{em:lineto}}
\put(  5.293,  6.471){\special{em:lineto}}
\put(  5.221,  6.394){\special{em:lineto}}
\put(  5.160,  6.348){\special{em:lineto}}
\put(  5.106,  6.353){\special{em:lineto}}
\put(  5.055,  6.411){\special{em:lineto}}
\put(  5.000,  6.500){\special{em:lineto}}
\put(  4.938,  6.587){\special{em:lineto}}
\put(  4.871,  6.638){\special{em:lineto}}
\put(  4.807,  6.631){\special{em:lineto}}
\put(  4.752,  6.569){\special{em:lineto}}
\put(  4.707,  6.471){\special{em:lineto}}
\put(  4.670,  6.373){\special{em:lineto}}
\put(  4.632,  6.306){\special{em:lineto}}
\put(  4.581,  6.291){\special{em:lineto}}
\put(  4.511,  6.325){\special{em:lineto}}
\put(  4.426,  6.386){\special{em:lineto}}
\put(  4.335,  6.442){\special{em:lineto}}
\put(  4.254,  6.464){\special{em:lineto}}
\put(  4.197,  6.433){\special{em:lineto}}
\put(  4.170,  6.354){\special{em:lineto}}
\put(  4.167,  6.247){\special{em:lineto}}
\put(  4.170,  6.142){\special{em:lineto}}
\put(  4.160,  6.066){\special{em:lineto}}
\put(  4.118,  6.032){\special{em:lineto}}
\put(  4.042,  6.037){\special{em:lineto}}
\put(  3.939,  6.061){\special{em:lineto}}
\put(  3.834,  6.078){\special{em:lineto}}
\put(  3.751,  6.067){\special{em:lineto}}
\put(  3.710,  6.017){\special{em:lineto}}
\put(  3.715,  5.934){\special{em:lineto}}
\put(  3.753,  5.833){\special{em:lineto}}
\put(  3.796,  5.738){\special{em:lineto}}
\put(  3.816,  5.663){\special{em:lineto}}
\put(  3.791,  5.616){\special{em:lineto}}
\put(  3.718,  5.591){\special{em:lineto}}
\put(  3.614,  5.574){\special{em:lineto}}
\put(  3.510,  5.550){\special{em:lineto}}
\put(  3.438,  5.508){\special{em:lineto}}
\put(  3.419,  5.446){\special{em:lineto}}
\put(  3.456,  5.371){\special{em:lineto}}
\put(  3.529,  5.293){\special{em:lineto}}
\put(  3.606,  5.221){\special{em:lineto}}
\put(  3.652,  5.160){\special{em:lineto}}
\put(  3.647,  5.106){\special{em:lineto}}
\put(  3.589,  5.055){\special{em:lineto}}
\put(  3.500,  5.000){\special{em:lineto}}
\end{picture}
\end{array}
\end{displaymath}
\caption{{\protect\small One loop corrections to $m_h^2$. The first
diagram gives a $\sim M_{Pl}^2$ contribution.}}
\end{center}
\label{higgscont}
\end{figure}
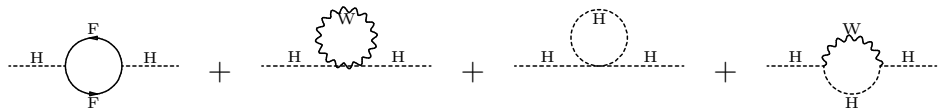
\newpage
\begin{flushleft}
{\bf Why SUSY?}
\end{flushleft}

Supersymmetry is an extra symmetry relating fermions and bosons, and so
provides some explanation of how particles of differing spin
should be related to one another. In an unbroken supersymmetric
theory, each fermion has a degenerate bosonic partner. Of course,
these so called superpartners are not observed, so that if
supersymmetry was ever the correct theory, it must have been broken.
However, with the introduction of superpartners at some rough energy
scale $M_{SUSY}$, the renormalisation group running of the gauge
couplings changes. The coupling constants are now seen to meet at a
scale $\sim 10^{16}$ GeV, as reflected by the correct $\sin^2
\theta_w$ prediction$^4$. $M_W$ becomes stabilised
because supersymmetry induces cancellations between the bosonic and
fermionic loop corrections to the mass. The quadratic divergences
induced by the loop corrections now add to zero and one is left with
merely logarithmic divergences.
\begin{flushleft}
\vspace{\baselineskip}
{\bf THE MINIMAL SUPERSYMMETRIC STANDARD MODEL (MSSM)}
\end{flushleft}

The MSSM is a minimal extension of the standard model into
supersymmetry.
In the model, every particle of the standard model has a superpartner
associated with it that transforms identically under the standard
model gauge group but have spin different by $\frac{1}{2}$. So for
example, each quark has a scalar "squark" superpartner, the gluons
have "gluinos" etc. At first sight however, the model has a U(1)$_Y^3$
gauge
anomaly.
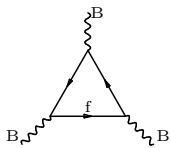
\begin{figure}
\begin{center}
\setlength{\unitlength}{0.25cm}
\begin{picture}( 10.0000, 10.0000)(   .000,   .000)
{\special{em:linewidth    .020cm}}
\put(  5.000,  4.000){\makebox(0,0){{\tiny f}}}
\put(  1.350,  2.180){\makebox(0,0)[br]{{\tiny B}}}
\put(  8.650,  2.180){\makebox(0,0)[bl]{{\tiny B}}}
\put(  5.150,  9.000){\makebox(0,0)[l]{{\tiny B}}}
\put(  5.000,  7.000){\special{em:moveto}}
\put(  3.000,  3.500){\special{em:lineto}}
\put(  4.175,  5.375){\special{em:moveto}}
\put(  3.903,  5.081){\special{em:lineto}}
\put(  4.019,  5.464){\special{em:lineto}}
\put(  4.175,  5.375){\special{em:lineto}}
\put(  3.000,  3.500){\special{em:moveto}}
\put(  7.000,  3.500){\special{em:lineto}}
\put(  4.805,  3.590){\special{em:moveto}}
\put(  5.195,  3.500){\special{em:lineto}}
\put(  4.805,  3.410){\special{em:lineto}}
\put(  4.805,  3.590){\special{em:lineto}}
\put(  7.000,  3.500){\special{em:moveto}}
\put(  5.000,  7.000){\special{em:lineto}}
\put(  6.019,  5.036){\special{em:moveto}}
\put(  5.903,  5.419){\special{em:lineto}}
\put(  6.175,  5.125){\special{em:lineto}}
\put(  6.019,  5.036){\special{em:lineto}}
\put(  3.000,  3.500){\special{em:moveto}}
\put(  3.025,  3.400){\special{em:lineto}}
\put(  3.026,  3.324){\special{em:lineto}}
\put(  2.988,  3.287){\special{em:lineto}}
\put(  2.912,  3.288){\special{em:lineto}}
\put(  2.813,  3.313){\special{em:lineto}}
\put(  2.713,  3.337){\special{em:lineto}}
\put(  2.637,  3.338){\special{em:lineto}}
\put(  2.599,  3.301){\special{em:lineto}}
\put(  2.600,  3.225){\special{em:lineto}}
\put(  2.625,  3.125){\special{em:lineto}}
\put(  2.650,  3.025){\special{em:lineto}}
\put(  2.651,  2.949){\special{em:lineto}}
\put(  2.613,  2.912){\special{em:lineto}}
\put(  2.537,  2.913){\special{em:lineto}}
\put(  2.438,  2.938){\special{em:lineto}}
\put(  2.338,  2.962){\special{em:lineto}}
\put(  2.262,  2.963){\special{em:lineto}}
\put(  2.224,  2.926){\special{em:lineto}}
\put(  2.225,  2.850){\special{em:lineto}}
\put(  2.250,  2.750){\special{em:lineto}}
\put(  2.275,  2.650){\special{em:lineto}}
\put(  2.276,  2.574){\special{em:lineto}}
\put(  2.238,  2.537){\special{em:lineto}}
\put(  2.162,  2.538){\special{em:lineto}}
\put(  2.063,  2.563){\special{em:lineto}}
\put(  1.963,  2.587){\special{em:lineto}}
\put(  1.887,  2.588){\special{em:lineto}}
\put(  1.849,  2.551){\special{em:lineto}}
\put(  1.850,  2.475){\special{em:lineto}}
\put(  1.875,  2.375){\special{em:lineto}}
\put(  1.900,  2.275){\special{em:lineto}}
\put(  1.901,  2.199){\special{em:lineto}}
\put(  1.863,  2.162){\special{em:lineto}}
\put(  1.787,  2.163){\special{em:lineto}}
\put(  1.688,  2.188){\special{em:lineto}}
\put(  1.588,  2.212){\special{em:lineto}}
\put(  1.512,  2.213){\special{em:lineto}}
\put(  1.474,  2.176){\special{em:lineto}}
\put(  1.475,  2.100){\special{em:lineto}}
\put(  1.500,  2.000){\special{em:lineto}}
\put(  5.000,  9.000){\special{em:moveto}}
\put(  5.088,  8.943){\special{em:lineto}}
\put(  5.143,  8.886){\special{em:lineto}}
\put(  5.143,  8.829){\special{em:lineto}}
\put(  5.088,  8.771){\special{em:lineto}}
\put(  5.000,  8.714){\special{em:lineto}}
\put(  4.912,  8.657){\special{em:lineto}}
\put(  4.857,  8.600){\special{em:lineto}}
\put(  4.857,  8.543){\special{em:lineto}}
\put(  4.912,  8.486){\special{em:lineto}}
\put(  5.000,  8.429){\special{em:lineto}}
\put(  5.088,  8.371){\special{em:lineto}}
\put(  5.143,  8.314){\special{em:lineto}}
\put(  5.143,  8.257){\special{em:lineto}}
\put(  5.088,  8.200){\special{em:lineto}}
\put(  5.000,  8.143){\special{em:lineto}}
\put(  4.912,  8.086){\special{em:lineto}}
\put(  4.857,  8.029){\special{em:lineto}}
\put(  4.857,  7.971){\special{em:lineto}}
\put(  4.912,  7.914){\special{em:lineto}}
\put(  5.000,  7.857){\special{em:lineto}}
\put(  5.088,  7.800){\special{em:lineto}}
\put(  5.143,  7.743){\special{em:lineto}}
\put(  5.143,  7.686){\special{em:lineto}}
\put(  5.088,  7.629){\special{em:lineto}}
\put(  5.000,  7.571){\special{em:lineto}}
\put(  4.912,  7.514){\special{em:lineto}}
\put(  4.857,  7.457){\special{em:lineto}}
\put(  4.857,  7.400){\special{em:lineto}}
\put(  4.912,  7.343){\special{em:lineto}}
\put(  5.000,  7.286){\special{em:lineto}}
\put(  5.088,  7.229){\special{em:lineto}}
\put(  5.143,  7.171){\special{em:lineto}}
\put(  5.143,  7.114){\special{em:lineto}}
\put(  5.088,  7.057){\special{em:lineto}}
\put(  5.000,  7.000){\special{em:lineto}}
\put(  7.000,  3.500){\special{em:moveto}}
\put(  7.100,  3.525){\special{em:lineto}}
\put(  7.176,  3.526){\special{em:lineto}}
\put(  7.213,  3.488){\special{em:lineto}}
\put(  7.212,  3.412){\special{em:lineto}}
\put(  7.188,  3.313){\special{em:lineto}}
\put(  7.163,  3.213){\special{em:lineto}}
\put(  7.162,  3.137){\special{em:lineto}}
\put(  7.199,  3.099){\special{em:lineto}}
\put(  7.275,  3.100){\special{em:lineto}}
\put(  7.375,  3.125){\special{em:lineto}}
\put(  7.475,  3.150){\special{em:lineto}}
\put(  7.551,  3.151){\special{em:lineto}}
\put(  7.588,  3.113){\special{em:lineto}}
\put(  7.587,  3.037){\special{em:lineto}}
\put(  7.563,  2.938){\special{em:lineto}}
\put(  7.538,  2.838){\special{em:lineto}}
\put(  7.537,  2.762){\special{em:lineto}}
\put(  7.574,  2.724){\special{em:lineto}}
\put(  7.650,  2.725){\special{em:lineto}}
\put(  7.750,  2.750){\special{em:lineto}}
\put(  7.850,  2.775){\special{em:lineto}}
\put(  7.926,  2.776){\special{em:lineto}}
\put(  7.963,  2.738){\special{em:lineto}}
\put(  7.962,  2.662){\special{em:lineto}}
\put(  7.938,  2.563){\special{em:lineto}}
\put(  7.913,  2.463){\special{em:lineto}}
\put(  7.912,  2.387){\special{em:lineto}}
\put(  7.949,  2.349){\special{em:lineto}}
\put(  8.025,  2.350){\special{em:lineto}}
\put(  8.125,  2.375){\special{em:lineto}}
\put(  8.225,  2.400){\special{em:lineto}}
\put(  8.301,  2.401){\special{em:lineto}}
\put(  8.338,  2.363){\special{em:lineto}}
\put(  8.337,  2.287){\special{em:lineto}}
\put(  8.313,  2.188){\special{em:lineto}}
\put(  8.288,  2.088){\special{em:lineto}}
\put(  8.287,  2.012){\special{em:lineto}}
\put(  8.324,  1.974){\special{em:lineto}}
\put(  8.400,  1.975){\special{em:lineto}}
\put(  8.500,  2.000){\special{em:lineto}}
\end{picture}
\caption{{\protect\small One loop triple hypercharge boson anomally.}}
\end{center}
\label{anomalli}
\end{figure}
This originates from the diagram with three B gauge
bosons connected to an  internal loop through which any fermions may
run (cf Fig.2)
and the counter term to it would destroy gauge invariance.
 The diagram is proportional to $\sum_i (Y_i/2)^3$ where i
runs over all active fermions. Through the hypercharge assignments,
this cancels in the standard model but in the MSSM the superpartner of
the Higgs called the Higgsino with $Y=1$ may run around the loop. To
cancel
this effect, a second Higgs $H_2$ must be introduced which transforms
in the same way to $H_1$ except for having $Y=-1$.

The new Higgs must also develop a vev $v_2$ to give masses to up
quarks and the two vevs are related by
\begin{equation}
\tan \beta = \frac{v_2}{v_1}
\end{equation}
where $v_1^2 + v_2^2 = v^2$ and $v=246$ GeV, the measured vev of
the standard model.

In chiral superfield form, the superpotential looks like
\begin{equation}
W_{MSSM} = U Q H_{2} u^c + D Q H_{1} d^c + E L H_1 e^c + \mu H_1 H_2
\label{superpot}
\end{equation}
where U, D and E are the up, down and charged lepton Yukawa matrices
respectively and all gauge and family indices have been suppressed.

One possible problem with this superpotential is the dimensionful
parameter $\mu$. $\mu$ needs to be $\sim M_Z$ to give
the right
electroweak symmetry breaking behaviour whereas one would expect it to
be of order of the new physics scale $M_{GUT}$. One solution to this
problem is described in the Next to Minimal Supersymmetric Standard
Model (NMSSM).
\begin{flushleft}
{\bf The NMSSM}
\end{flushleft}

The $\mu$ term in Eq.\ref{superpot} is replaced by $\lambda N H_1 H_2$
where N is a gauge singlet and therefore doesn't affect the coupling
constant unification. In certain supergravity models, N develops a vev
naturally of order $M_Z$ and so the $\mu$ term is generated without
having to put $\mu$ in "by hand." The superpotential now has a
discrete Peccei-Quinn symmetry which leads to phenomenologically
unacceptable low energy axions and so a term $-\frac{k}{3} N^3$ is
added which breaks it. \footnote{ $\lambda$ and $k$ are merely
coupling constants}
\begin{flushleft}
\vspace{\baselineskip}
{\bf GUTS WITH YUKAWA UNIFICATION}
\end{flushleft}

GUTs can quite naturally provide Yukawa unification relations between the
quarks and/or leptons. For example in SU(5), the right handed down
quarks and conjugated lepton doublet lie in a \underline{5}
representation. When a mass term $\sim 5^i 5_i$ is formed, the Yukawa
relation
\begin{equation}
\lambda_b ( M_{GUT} )= \lambda_\tau (M_{GUT})
\label{unification}
\end{equation}
applies. Also in SO(10), the whole of one family and a right handed
neutrino is contained in one \underline{16} representation, leading to
triple Yukawa unification, where the top, bottom and charged lepton
Yukawa couplings are equal at the GUT scale.

These relations can be used to constrain the parameter space of $m_t$
and $\tan \beta$, which has been done for the MSSM$^5$. Our
idea was to repeat this calculation for the NMSSM, to see how much the
viable parameter space changes in the model.
\begin{flushleft}
\vspace{\baselineskip}
{\bf THE CALCULATION}
\end{flushleft}

The basic idea is to choose some $\tan \beta$ and $m_t$ and run
$\lambda_b$ and $\lambda_\tau$ up to $M_{GUT} \sim 10^{16}$ GeV. Then,
to some arbitrary accuracy, one can determine whether the GUT relation
Eq.\ref{unification} holds. If it does, then SU(5) and the other Yukawa
unifying extensions of the standard model are possible on this point
in parameter space. The procedure is iterated over all reasonable
values of $\tan \beta$ and $m_t$. The calculation is presented in
more detail in Ref.6.
\newpage
\begin{flushleft}
{\bf Starting Point $M_Z$.}
\end{flushleft}

We use the definitions of the gauge couplings at $M_Z$: $\alpha_1^{-1}
(M_Z)=58.89$, $\alpha_2^{-1} (M_Z)=29.75$ and $\alpha_3^{-1}(M_Z)=0.11
\pm 0.01$. The first two gauge couplings are determined accurately
enough for our purposes whereas the third needs to be used as a
parameter, on account of its large uncertainty.

In order to convert masses of quarks to Yukawa couplings, we simply
need to read them off the potential Eq.\ref{superpot} at some energy
scale (taken here to be $m_t$):
\begin{eqnarray}
\lambda_{t} \left( m_{t} \right) & = & \frac{\sqrt{2} m_{t}\left(
m_{t}
\right) }{v \sin\beta}\\
\lambda_{b} \left( m_{t} \right) & = & \frac{\sqrt{2} m_{b}\left(
m_{b}
\right)}{\eta_{b} v \cos \beta }\\
\lambda_{\tau} \left( m_{t} \right) & = & \frac{\sqrt{2} m_{\tau}
\left(
m_{\tau} \right)}{\eta_{\tau} v \cos\beta}. \label{Yuk}
\end{eqnarray}
 where
\begin{equation}
 \eta_{f} =\frac{m_{f} \left( m_{f} \right)}{m_{f} \left( m_{t}
\right)}.
\end{equation}
 Note that whereas the $m_t$ referred to here is always the
running one,
it can be related to the physical mass by$^5$
\begin{equation}
m_t^{phys} = m_t \left( m_t \right) \left[ 1 + \frac{4}{3 \pi}
\alpha_3 \left( m_t \right) + O \left( \alpha_3^2 \right) \right].
\end{equation}

To determine $\eta_b$ and $\eta_\tau$, the masses are run up from the
on shell mass to $m_t$ using effective 3~loop~QCD~$\otimes$~1~loop~QED
$^{7,8,9,10}$.
Note that these factors will depend of
$m_b=4.25 \pm 0.15$ GeV and $\alpha_3 (M_Z)$.
\begin{figure}
\begin{center}
\leavevmode
\hbox{%
\epsfxsize=4.5in
\epsfysize=2.6in
\epsffile{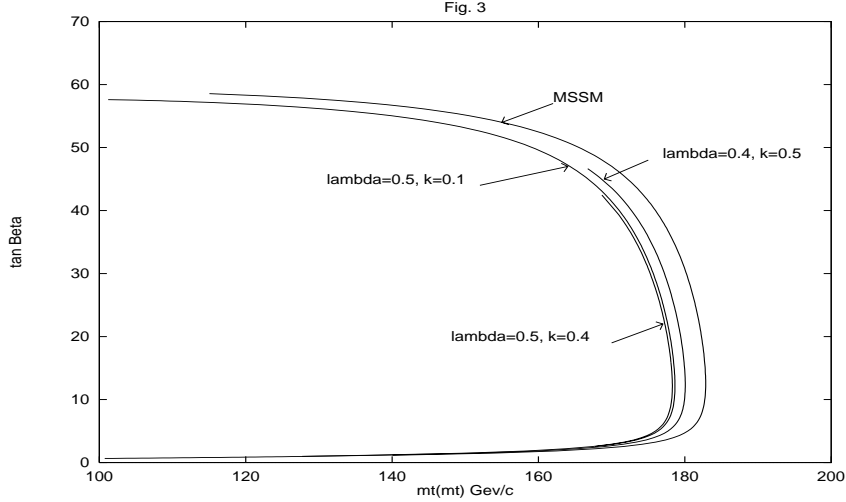}}
\end{center}
\caption{{\protect\small
Viable Range of Parameter Space For $\alpha_S (M_Z)=0.11$,
$m_b=4.25$ GeV. $\lambda$ and $k$ values are quoted at $m_t$.}}
\label{prm}
\end{figure}
$m_t$ is assumed to be the rough energy scale when the whole
supersymmetric spectrum kicks in. While being unrealistic, trials with
$M_{SUSY} = 1$ TeV show only a few percent deviation from the
predictions with $M_{SUSY} = m_t$. So, having determined the gauge and
relevant Yukawa couplings at $m_t$, we need RG equations to run them
up to $M_{GUT}$ in the NMSSM. To derive these, we used results from a
general superpotential$^{11}$ to obtain
\begin{eqnarray}
16 \pi^{2} \frac{\partial \lambda_{t}}{\partial t} &=& \lambda_{t}
\left[ 6\lambda_{t}^{2} +  \lambda_{b}^{2} +  \lambda^{2} -
\left( \frac{13}{15}
g_{1}^{2} + 3g_{2}^{2} + \frac{16}{3}g_{3}^{2} \right)
\right] \nonumber \\
16 \pi^{2} \frac{\partial \lambda_{b}}{\partial t} &=&
\lambda_{b} \left[ 6
\lambda_{b}^{2} + \lambda_{\tau}^{2} + \lambda_{t}^{2} + \lambda^{2} -
\left(
\frac{7}{15} g_{1}^{2} + 3g_{2}^{2} +
\frac{16}{3} g_{3}^{2} \right) \right] \nonumber \\
16 \pi^{2} \frac{\partial \lambda_{\tau}}{\partial t} &=&
\lambda_{\tau}
\left[ \lambda_{\tau}^{2} + 3 \lambda_{b}^{2}  + \lambda^{2}
- \left( \frac{9}{5} g_{1}^{2}
+ 3g_{2}^{2} \right) \right] \nonumber \\
16 \pi^{2} \frac{\partial \lambda}{\partial t} &=& \lambda
\left[ 4 \lambda^{2} + 2 k^{2} + 3\lambda_{\tau}^{2} + 3 \lambda_{b}^{2}
+ 3 \lambda_{t}^{2}
- \left(
\frac{3}{5} g_{1}^{2} + 3g_{2}^{2} \right) \right] \nonumber \\
16 \pi^{2} \frac{\partial k}{\partial t} &=& 6 k \left[ \lambda^{2}+
k^{2}
\right] \label{krg}
\end{eqnarray}
in the limit that the lighter two families have negligible
contributions (a very good approximation).

The Yukawa couplings can now be run from $m_t$ to $10^{16}$ GeV
using numerical techniques. The parameters $\lambda$ and $k$ particular
to the NMSSM are unconstrained at $m_t$ so they are merely varied for
different curves.

Our results are displayed in Fig. \ref{prm} as
contours in the $\tan \beta - m_t$ plane consistent
with Eq.\ref{unification}. We take
$\alpha_3(M_Z)=0.11$, $m_b=4.25 GeV$ and the NMSSM parameters
$\lambda (m_t)$ and $k(m_t)$ as indicated.
The MSSM contour is shown for comparison and is
indistinguishable from the NMSSM contour with
$\lambda \left( m_{t} \right) =0.1$ and $k \left( m_{t} \right) =0.5$.
In fact our plot for the MSSM based on 1-loop RG equations
is very similar to the 2-loop result in ref.5.
The deviation of the NMSSM contours from the MSSM contour
depends most sensitively on $\lambda(m_t)$ rather than $k(m_t)$.
Two of the contours are shortened due to either $\lambda$ or $k$
blowing up at the GUT scale.
For $\lambda \left(
m_t \right) =0.5$, $k \left( m_t \right) =0.5$, no points in
the $m_{t} - \tan \beta $ plane are consistent with Eq.\ref{unification}
Yukawa unification, while for $\lambda \left(
m_t \right) =0.1$, $k \left( m_t \right) =0.1-0.5$
the contours are virtually indistinguishable from the MSSM contour.
In general we find that for
any of the current experimental limits on $\alpha_3$ and $m_b$, the
maximum value of $\lambda (m_t)$ or $k (m_t)$ is $\sim 0.7$ for a
perturbative solution to Eq.\ref{unification}.
\begin{figure}
\begin{center}
\leavevmode
\hbox{%
\epsfxsize=4.5in
\epsfysize=2.6in
\epsffile{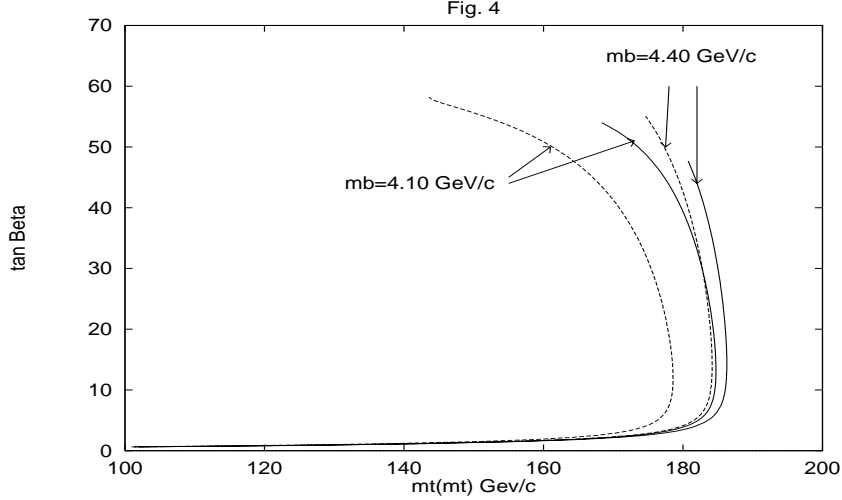}}
\end{center}
\caption{{\protect\small
Viable Range of Parameter Space For $\alpha_S (M_Z)=0.12$
and experimental bounds of $m_b=4.1$--$4.4$ GeV. The left most lines
are for $\lambda_b=0.9 \lambda_\tau$.}}
\label{thresh}
\end{figure}

Fig.\ref{thresh} shows the effects of particle thresholds, which
can modify
Eq.\ref{unification} to $\lambda_b = 0.9 \lambda_\tau$. Our treatment
does not treat supersymmetric or heavy thresholds exactly and so some
sort of corrections like those shown are expected. The curves are at
$\alpha_S (M_Z) =0.12$ and $m_b=4.1$--$4.4$ GeV to illustrate that
uncertainties in these quantities make a large difference to the
parameter space. These uncertainties are much bigger than those
associated with the NMSSM, and so the MSSM and NMSSM would be
practically indistinguishable given the parameters $m_t$ and $\tan
\beta$.
\begin{flushleft}
{\bf Other Yukawa Parameters}
\end{flushleft}

The next useful step is to notice that Eqs.\ref{krg} are all of the form
\begin{equation}
16 \pi^2 \frac{\partial \lambda_a}{\partial t} = \lambda_a \left[
\sum_i M_i^a \lambda_i^2 - \sum_{j=1}^3 c_j^a g_j^2 \right],\\
\end{equation}
where $M_i^a$ and $c_j^a$ are constants supplied by the relevant RG
equation.
When the $\beta$ function
\begin{equation}
\frac{d g_i}{d t} = \frac{b_i g_i^3}{16 \pi^2}
\end{equation}
is inserted, and the RG equations are reparameterised in terms of the
flow and not the trajectory of the solutions, we obtain
\begin{equation}
\frac{\lambda_a (M_{SUSY})}{\lambda_a (M_{GUT})} = \xi^a \mbox{exp} (
-\sum_i M_i^a I_i),
\end{equation}
where
\begin{equation}
\xi^a = \prod_{i=1}^3 \left( \frac{\alpha (M_{GUT})}{\alpha_i(M_{SUSY})}
\right)^\frac{c_i^a}{2 b_i}
\end{equation}
contains all the information about the gauge couplings and
\begin{equation}
I_i =
\frac{1}{16 \pi^2} \int^{\ln (M_{GUT})}_{\ln (M_{SUSY}) }
\lambda_i^2 \mbox{d} t \label{iint}
\end{equation}
concerns the Yukawa couplings.

With this formulation, the running of the physically
relevant Yukawa eigenvalues and mixing angles can be expressed in
simple terms as shown below,
\begin{eqnarray}
\left( \frac{\lambda_{u,c}}{\lambda_{t}} \right)_{M_{SUSY}} & = &
\left( \frac{\lambda_{u,c}}
{\lambda_{t}} \right)_{M_{GUT}} e^{3I_{t} + I_{b}} \label{start}
\nonumber \\
\left( \frac{\lambda_{d,s}}{\lambda_{b}} \right)_{M_{SUSY}} & = &
\left( \frac{\lambda_{d,s}}
{\lambda_{b}} \right)_{M_{GUT}} e^{3I_{b} + I_{t}} \nonumber \\
\left( \frac{\lambda_{e, \mu}}{\lambda_{\tau}} \right)_{M_{SUSY}} & =
& \left(
\frac{\lambda_{e, \mu}}
{\lambda_{\tau}} \right)_{M_{GUT}} e^{3I_{\tau}} \nonumber \\
\frac{ \mid V_{cb} \mid _{M_{GUT}}}{\mid V_{cb} \mid
_{M_{SUSY}} }
& = & e^{I_{b}+I_{t}} , \label{GUTsusy}
\end{eqnarray}
with identical scaling behaviour to $V_{cb}$ of $V_{ub}$,
$V_{ts}$, $V_{td}$.
To a consistent level of approximation $V_{us}$, $V_{ud}$, $V_{cs}$,
$V_{cd}$, $V_{tb}$, $\lambda_{u}$/$\lambda_{c}$,
$\lambda_{d}$/$\lambda_{s}$ and $\lambda_{e}$/$\lambda_{\mu}$ are RG
invariant.
The CP violating quantity J scales as $V_{cb}^{2}$.
Eqs. \ref{GUTsusy}, \ref{iint} also apply to the NMSSM since
the extra $\lambda$ and $k$ parameters
cancel out of the RG equations in a similar way to the gauge contributions
as can easily be seen from Eq.\ref{krg}.
The only difference to these physically relevant
quantities is therefore contained in $I_{\tau}$, $I_{b}$ and $I_{t}$.

\begin{figure}
\begin{center}
\leavevmode
\hbox{%
\epsfxsize=4.5in
\epsfysize=2.6in
\epsffile{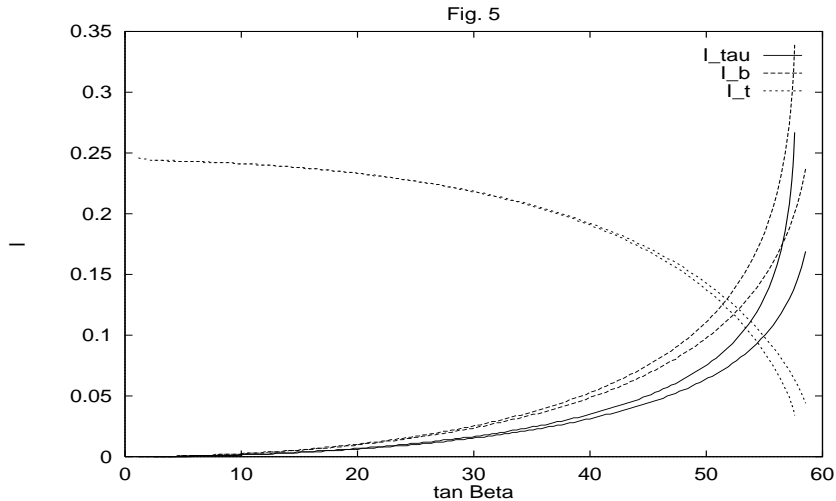}}
\end{center}
\caption{{\protect\small$I_i$ integrals for $\alpha_S (M_Z)=0.11$,~
$m_b=4.25$ GeV. }}
\label{Is}
\end{figure}

These $I_i$ integrals are shown in Fig.\ref{Is} and the NMSSM results
are the upper
lines of each pair, and it is clear that the deviation between the two
models is small again.

We emphasise that the results of the $I_i$ integrals
shown in Fig.\ref{Is} play a key role in determining
the entire fermion mass spectrum via the scaling relations
of Eq.\ref{GUTsusy}. The small deviation between the NMSSM and the MSSM
results compared to the experimental uncertainties
means that the recent GUT scale
texture analyses of the quark mass matrices which were performed
for the MSSM are equally applicable to the NMSSM.
For example, the recent Ramond, Roberts and Ross (RRR)$^{12}$
texture analysis is also based upon Eq.\ref{unification} and
assumes a Georgi-Jarlskog
(GJ)$^{13,14}$ ansatze
for the charged lepton Yukawa matrices, although their results in the
quark sector
are insensitive to the lepton sector. It is clear that all the RRR
results are immediately applicable to the NMSSM
since the only difference between the two models
enters through the scaling integrals $I_i$ whose deviation
we have shown to be
negligible compared to the experimental errors.
\begin{flushleft}
\vspace{\baselineskip}
{\bf CONCLUSIONS}
\end{flushleft}

We have discussed the unification of the bottom quark
and tau lepton Yukawa couplings within the framework of the
NMSSM. By comparing the allowed regions of the $m_t$-$\tan \beta$ plane
to those in the MSSM
we find that over much of the parameter space
the deviation between the predictions of the two models
which is controlled by the parameter $\lambda$ is small,
and always much less than the effect of current
theoretical and experimental uncertainties in
the bottom quark mass and the strong coupling constant.
We have also discussed the scaling of the light fermion masses and
mixing angles, and shown that to within current uncertainties,
the results of recent quark texture analyses$^{12}$
performed for the minimal
model also apply to the next-to-minimal model. There are however
two distinguishing features of the NMSSM. Firstly, the scaling
of the charged lepton masses will be somewhat different,
depending on $\lambda$ and $k$. Although this will not affect
the quark texture analysis of RRR, it may affect the success of the
GJ ansatze$^{13,14}$ for example. Secondly, the larger
$\tan \beta$ regions may not be accessible in the NMSSM
for large values of $\lambda$ and $k$, so that full Yukawa unification
may not be possible in this case.
\newpage
\begin{flushleft}
{\bf REFERENCES}
\vspace{\baselineskip}

\begin{tabular}{rp{14.5cm}}
$[1]$& V.~Barger and R.~J.~N. Phillips,
 Preprint MAD/PH/752  (1993).\\
$[2]$& M.~Chanowitz, J.~Ellis, and M.~K. Gaillard,
 Nuclear Physics B128, 506 (1977).\\
$[3]$&A.~J. Buras, J.~Ellis, M.~K. Gaillard, and D.~V. Nanopoulos,
 Nucl. Phys. B135, 66 (1978).\\
$[4]$& H.~E. Haber,
 Preprint SCIPP 92/33  (1993).\\
$[5]$& V.Barger, M.S.Berger, and P.Ohmann,
 Phys. Rev. D47, 1093 (1993).\\
$[6]$& B.~C. Allanach and S.~F. King
 Phys. Lett. B328, 360 (1994).\\
$[7]$& S.G.Gorishny, A.L.Kataev, S.A.Larin, and L.R.Surgaladze,
 Mod. Phys. Lett. A5, 2703 (1990).\\
$[8]$& O.V.Tarasov, A.A.Vladimirov, and A.Yu.Zharkov,
 Phys. Lett. B93, 429 (1980).\\
$[9]$& S.G.Korishny, A.L.Kataev, S.A.Larin, and P.~Lett.,
 Phys. Lett. B135, 457 (1984).\\
$[10]$& L.Hall,
 Nucl. Phys. B75 (1981).\\
$[11]$& S.P.Martin and M.T.Vaughn,
 NUB-3081-93TH hep-ph 9311340 (1993).\\
$[12]$& P.~Ramond, R.~Roberts, and G.~Ross,
 RAL-93-010 UFIFT-93-06 (1993).\\
$[13]$& H.~Georgi and C.~Jarlskog,
 Phys. Lett. B86, 297 (1979).\\
$[14]$& S.~Dimopoulos, L.~Hall, and S.~Raby,
 Phys. Rev. D45, 4192 (1992).
\end{tabular}
\end{flushleft}
\end{document}